# Spectral broadening of 2-mJ femtosecond pulses in a compact air-filled convex-concave multi-pass cell

ALAN OMAR,* TIM VOGEL, MARTIN HOFFMANN, AND CLARA J. SARACENO

*Photonics and Ultrafast Laser Science, Ruhr-Universität Bochum, Universitätsstraße 150, 44801 Bochum, Germany*
*\*Corresponding author: [alan.omar@ruhr-uni-bochum.de](mailto:alan.omar@ruhr-uni-bochum.de)*

**Multi-pass cell (MPC) based temporal pulse compressors have emerged in the last years as a powerful and versatile solution to the intrinsic issue of long pulses from Yb-based high-power ultrafast lasers. However, the spectral broadening of high-energy (typically more than 100 µJ) pulses has only been realized in complex setups, i.e., in large and costly, pressure-controlled vacuum chambers to avoid strong focusing, ionization, and damage on the mirrors. Here, we present spectral broadening of 2-mJ pulses in a simple and compact (60-cm long) multi-pass cell operated in ambient air. Instead of the traditional Herriott cell with concave-concave (CC/CC) mirrors, we use a convex-concave (CX/CC) design, where the beam stays large at all times allowing both to minimize damage and operate in ambient air. We demonstrate spectral broadening of 2.1-mJ pulses at 100-kHz repetition rate (200-W of average power) from 2.1 nm (pulse duration of 670 fs) to a spectral bandwidth of 24.5 nm, supporting 133-fs pulses with 96% transmission efficiency. We show the compressibility of these pulses down to 134 fs, and verify that the spectral homogeneity of the beam is similar to previously reported CC/CC designs. To the best of our knowledge, this is the first report of a CX/CC MPC compressor, operated at high pulse energies in air. Because of its simplicity, small footprint and low cost, we believe this demonstration will have significant impact in the ultrafast laser community.**

## 1. Introduction

High-average power ultrafast lasers at 1-µm wavelength based on Ytterbium-doped gain media have seen tremendous progress in the last decade [1–3], with average power levels largely exceeding the kilowatt and increasingly widespread commercial availability. However, comparatively to this progress, their adoption in many applications and particularly in scientific research has been slow, due to the lack of efficient and simple paths to temporally reduce the typically long pulse durations of >300 fs to the realm of Ti:Sapphire lasers, which commonly operate well-below 100 fs, even with multi-millijoule pulse energies.

For many years, gas-filled hollow-core capillaries [4] or filament compressors [5] were the mainstream techniques for nonlinear spectral broadening of millijoule class pulses via self-phase modulation (SPM) in noble gases and subsequent pulse compression using dispersive delay lines. Using this technique, both high-average power and mJ-energies have been successfully compressed down to few-cycle pulses. Some recent state-of-the-art demonstrations of the before mentioned techniques are a two stage compression with gas-filled capillaries [6], and compression with a self-trapping filament in air [7]. In the first example, 240-fs pulses at 660 W of average power were compressed to 6.3 fs and 216 W with an optical transmission of 32% using Ne gas in the second stage at 7.5 bar. In the second example, 200-fs pulses were compressed at 4 mJ in a self-channeling region of 4.6 m to 40 fs. However, these techniques suffer from low efficiency, typically in the 50% range, and their practical implementation is cumbersome due to the complexity of the compression setup.

Since their first demonstration in 2016 [8], Herriott-type multi-pass cells (MPC) have been recognized as a powerful technique to spectrally-broaden and compress pulses with extremely high throughput of >90% and in a very wide range of input parameters and wavelength regions [9–11]. The main operation principle of spectral-broadening in MPCs is to avoid beam degradation and catastrophic beam collapse by dividing the nonlinear spectral broadening into small enough steps with sufficient free-space propagation in between, which suppresses spatio-temporal coupling [12]. This combines the advantages of free-space propagation – which is critical for high average power lasers – while preserving an excellent beam quality even with large broadening factors. In this goal, the concave-concave CC/CC Herriott-type MPC design is practical to introduce these multiple steps in a compact footprint with small misalignment sensitivity [13]. The nonlinear medium used to obtain spectral broadening can be a solid-state material (for example fused silica) for low energy systems (typically <100 µJ), or the gas itself inside the cell for higher pulse energy systems, with different scaling and design laws. So far, only the CC/CC Herriott-cell design has been used for pulse energies of >100 µJ using noble gases such as Ar, Kr, or Xe as a nonlinear medium. In this way, remarkable results have been achieved; for instance, pulses from a 1-kW Yb:fiber laser system with a pulse duration of 200 fs were compressed to 31 fs at 1 mJ and an optical transmission of 96% in an argon-filled MPC in a 1.5-m long chamber [14]. A similar technique was used by Kaumanns et al. to compress pulses with 18 mJ of pulse energy from 1.3 ps to 41 fs in a low-pressure chamber with a length of 3 m [15]. Balla et al. compressed 1.2-ps and 2-mJ pulses to a pulse duration of 32 fs with 1.6 mJ in a 2.5-m long Kr-filled MPC with optical transmission of 80% [16]. Recently, Pfaff et al. spectrally broadened 840-fs, 10-mJ pulses and showed compressibility to 38 fs, also with high optical transmission of 97% with a low-pressure chamber length of 2.7 m [17]. While these results are certainly very impressive, they can only be achieved in a CC/CC MPC if critical parameters are carefully controlled.

In scaling the energy in the MPC, the naturally occurring focus in the CC/CC MPC has to be kept reasonably large. Otherwise, one can easily exceed the maximum nonlinear phase shift per pass that avoids spatio-temporal coupling, or the high peak intensity can give rise to gas ionization and subsequent beam degradation and pulse distortion. So far, this issue has been addressed by scaling the length of the gas chamber and the radii of curvature of the MPC mirrors, leading to impractical and expensive compression setups of several meters in length [18]. The idea of using a CX/CC design to reduce footprint and scale the pulse energy has been discussed in [19], but to the best of our knowledge not been demonstrated so far.

Here, we present the use of a compact convex-concave (CX/CC) MPC in ambient air, i.e. without any vacuum or high-pressure chamber at 2.1 mJ pulse energy. This was possible with a carefully tailored MPC design using fast 3D pulse propagation simulations. As a first proof-of-principle result, we show nonlinear spectral broadening of 2.1-mJ, 670-fs pulses at 210 W of average power from 2.1 nm to 24.5 nm corresponding to a transform limit (TL) of 133 fs. We show compressibility of the pulses down to 134 fs with comparable homogeneity to previously demonstrated designs at this high pulse energy. Furthermore, our air-filled cell is 60-cm long, and only requires standard highly-reflective mirrors for spectral broadening. We believe this is an important demonstration toward more widely accessible Yb-doped high-power systems with short pulses. Our experiment was limited by the damage threshold of the available optics at the time of the experiment, and we believe further scaling to much higher energies should be straightforward to implement.

## 2. Experimental setup and results

The driving laser system is a commercially available Yb-doped thin-disk regenerative amplifier (Trumpf Dira 500-10) with a maximum average power of 500 W, operated at a repetition rate of 100 kHz, corresponding to a pulse energy of 5 mJ, and a constant pulse duration of 670 fs with a spectral width of 2.1 nm. The beam profile quality $M^2$ at the laser's output is 1.25x1.28 (In $X$ and $Y$) with a slightly asymmetric profile. Using these parameters as input, we studied possible designs of CX/CC cells using numerical simulations. Driven by constraints in the available choice of CX mirrors in our laboratory at the time of the experiment (radius of curvature: -2.5 m), we studied an air-based MPC compressor using our home-made 3D pulse propagation tool.

Our numerical model solves the spatial-temporal nonlinear Schrödinger equation in three dimensions ($x$, $y$, $t$), using the split-step Fourier method and takes into account linear effects (diffraction and dispersion) and nonlinear effects (optical Kerr effect, self-steepening and instantaneous and delayed Raman scattering). The delayed Raman response is taken into account in air at room temperature (295 K) and in this model, air is considered as a mixture of 80% of $N_2$ and 20% of $O_2$ at atmospheric pressure [20]. In order to optimize the design and access a wide range of parameters, the code is written in C++ and Radeon Open Compute (ROCM) and is executed on a graphics processing unit (GPU) to reduce the computational time of one simulation. Moreover, the nonlinear refractive index of air in our simulation is 2.6 × 10$^{-23}$ m$^2$/W as measured by J. Schwarz et al. at a wavelength of 1054 nm and pulse duration of 540 fs using a wavefront sensor technique [21].

For the design of our CX/CC MPCs we used a 2-inch plano-CC mirror with a ROC of 2 m, and for the second mirror a 2-inch plano-CX mirror with a ROC is -2.5 m. The mirrors were separated by 61.5 cm, and we used 15 full roundtrips, corresponding to 30 reflections on either of the MPC mirrors. Both mirrors are high reflectors with close to zero group delay dispersion (GDD). The group velocity dispersion (GVD) of air at 1030 nm is 0.0162 fs$^2$/mm, which in our case results in a GDD per pass of 10 fs$^2$, and can therefore be neglected.

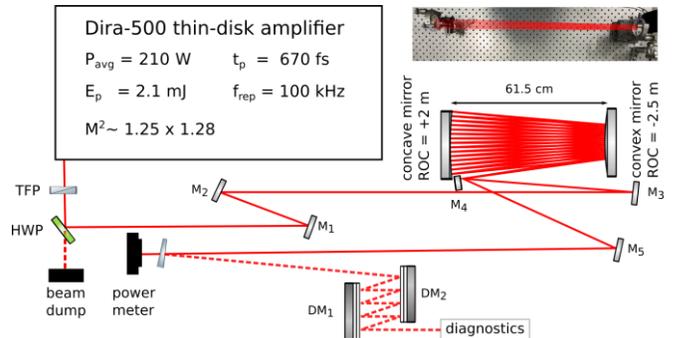

**Fig. 1.** Schematic of the setup. An Yb-doped thin-disk regenerative amplifier seeds the asymmetric air-filled CX/CC MPC. HWP: Zero-order half-wave plate, TFP: Thin-film polarizer, M1 and M2: Curved HR mirrors are used for high power mode matching, M4: in-coupling mirror, M5: recollimation mirror, PM: power meter, DM: dispersive mirrors.

In Fig. 1, the schematic of the experimental setup is shown. The beam of the amplifier is attenuated by a half-wave plate (HWP) and a thin-film polarizer (TFP). Two spherical mirrors are used to adjust the mode of the input beam to the necessary beam caustic as simulated using our 3D code, that allows for a controlled propagation through the MPC without reaching the damage threshold of the MPC mirrors. The beam then propagates through the MPC in a closed-path configuration where the beam is coupled in and out via a rectangular mirror. The nonlinear spectral broadening occurs in air with an estimated average nonlinear phase of 0.15π per pass. The estimated beam size (1/e$^2$) on the 7-mm rectangular in-coupling mirror is 1.65 mm which limits the number of spots on the 2-inch CC mirror in our experiment, consequently preventing more nonlinear spectral broadening by increasing the number of passes. In addition, achieving more spectral broadening by increasing the pulse energy is limited in this design by the damage threshold of the CX mirror available at the time of the experiment, which is observed at a fluence of 0.3 J/cm$^2$. We are therefore in this specific experiment limited to a spectral broadening of a factor of approximately 10 (from 2.1 nm to 24.5 nm); however future experiments with other adapted optical elements should allow to reach shorter pulses. We note that in spite of being operated in air, our simulations do not predict a significant influence of Raman scattering down to pulse durations of 50 fs.

Even though we have high average power, we have not observed a significant temperature rise on the mirror coating or the optomechanical elements, measured by a thermal camera. The MPC setup is placed in a housing to reduce the beam fluctuations due to air turbulence, however, the housing is in this case not sealed. The output beam is collimated with a spherical mirror, and the maximum output power is 201 W, which results in a power transmission efficiency of 96%. In our cell, the peak power is slightly below the critical peak power for self-focusing in air [21], as required in gas-filled MPCs [22]. For significantly higher peak powers a slight low-pressure, but still very compact, air-filled chamber can be easily implemented.

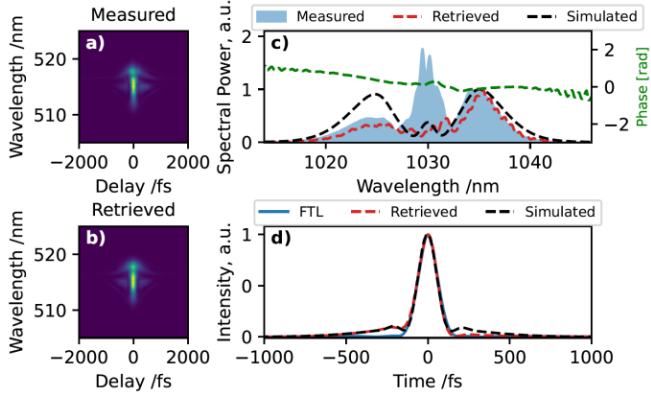

**Fig. 2.** Compression results measured with SHG-FROG. a) and b) Measured and retrieved traces. c) Measured, retrieved, and simulated spectrum and spectral phase of the compressed pulse. d) FTL, retrieved and simulated temporal profile of the pulse. The blue curve is the transform limit of the measured spectrum indicating a pulse duration of 133 fs.

The reflections on the MPC mirrors are the main cause of residual power loss observed. At the time of the experiments, we did not have large dispersion optics to enlarge the beam sufficiently, therefore, to prevent damage on the dispersive mirrors, the power is attenuated with two silica wedges and compressed with 38 reflections on dispersive mirrors with a total GDD of -18 kfs$^2$. The same compressibility is expected from the high-power beam given the availability of large enough dispersive optics. We characterized the compressed pulses using a home-built second-harmonic generation (SHG) frequency-resolved optical gating (FROG) setup. Fig. 2 a) and b) show the results of the measured and retrieved traces of the compressed pulse. Both traces are in good agreement with each other, with a FROG error of 0.5% using a 512x512 grid. The blue-filled curve in Fig. 2 c) shows the measured spectrum after the compression stage using an optical spectrum analyzer (OSA) with good agreement with the retrieved (dashed red) and simulated spectrum (dashed black), and the green curve is the retrieved spectral phase of the compressed pulse.

The central peak around 1030 nm in the measured spectrum relates to the amplifier spectrum and the characteristic pre- and post-pulses of the regenerative amplifier due to the leakage of the Pockels cell. These pulses do not contribute to the spectral broadening due to the extremely low energy in a single side pulse, and the estimated deposited energy in these side pulses is 6%. In addition, amplified spontaneous emission (ASE) contributes approximately 1% of the energy deposited in the central lobe [14,17]. The calculated spectral bandwidth is 24.5 nm, corresponding to the Fourier-transform limit (FTL) of 133 fs. In Fig. 2 d), the retrieved temporal intensity profile (dashed red) is nearly transform-limited (blue) with good agreement with the simulated pulse (dashed black). It exhibits a pulse duration at full width at half-maximum (FWHM) of 134 fs with a compression factor of 5.

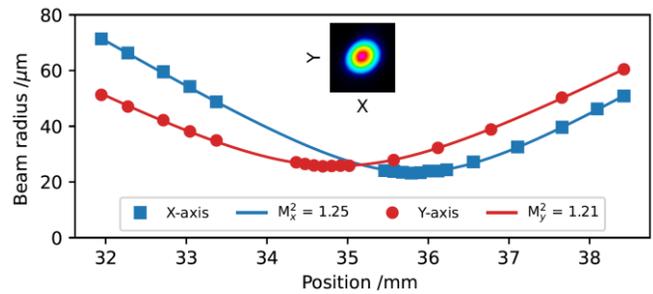

**Fig. 3.** Beam quality factor measurement ($M^2$) of the output beam along the transverse axes. The inset shows an image of the beam profile, measured by a CMOS camera.

Fig. 3 depicts the good beam quality of the compressed pulse with a measured $M_x^2 \times M_y^2$ = 1.21 x 1.25, which is comparable to the laser's output beam profile quality $M^2$ of 1.25 x 1.28. To investigate the spatio-spectral homogeneity of this configuration, we characterize the spectral homogeneity of the compressed beam by measuring the $V$-parameter [12]. 40 spectra are measured along the $x$-axis and $y$-axis of the collimated beam of a beam diameter (1/e$^2$) of $2w_x$ x $2w_y$ = 2.6 x 2.7 mm by using a multimode fiber with a mode size of 200 µm (SMA connector) coupled into a spectrometer (Avantes Starline). The fiber is positioned on an $XY$-motorized stage for repeatable and accurate measurements with a step size of 150 µm. Fig. 4 a) and b) depict the spectral homogeneity of both axes. In Fig 4. c) and d) the $V$-parameter (solid red) is over 82% ($x$-axis) and 81% ($y$-axis) within a 1/e$^2$ beam area. The intensity-weighted average $V$-parameter is ~90% for both axes within a 5-mm diameter of the measurement area. These results match the homogeneity measurements of other gas-filled Herriott-type MPCs, such as $V$-parameter measured by Pfaff et al [17].

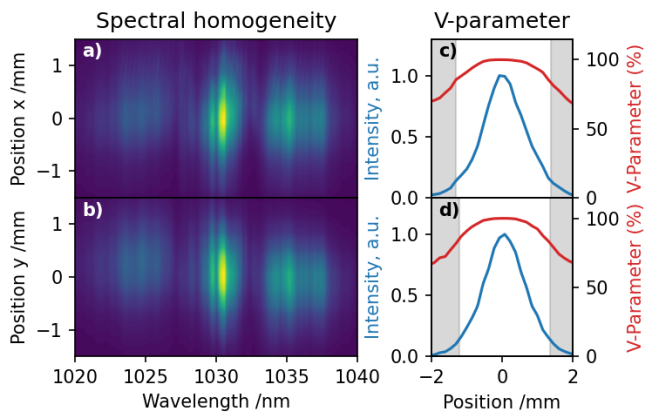

**Fig 4.** Spatial-spectral homogeneity characterization of the compressed pulse for both axes. The solid blue curves on the right show the normalized intensity of the spectral profile. The red solid curves show the corresponding spatial-spectral homogeneity values (*V*-parameter), and the grey boxes show the beam area within a beam diameter ($1/e^2$).

## 3. Conclusion and outlook

In conclusion, we present for the first time to the best of our knowledge the use of a convex-concave multi-pass cell in air for spectral broadening of mJ pulses. We spectrally broadened the output pulses of a thin-disk regenerative amplifier from 670 fs, 2.1 mJ at 1030 nm, at a high average power of 210 W (100 kHz repetition rate) to 134 fs and 203 W with an excellent optical efficiency of 96%, preserving the beam quality $M^2 < 1.3$ and obtaining a good spectral homogeneity. This novel scheme of asymmetric MPC represents a very compact and simple compressor to bring high-power and high-energy ultrafast lasers to sub-100 fs and beyond. Compared to the more standard utilized CC/CC MPC, it has the advantage of being much more compact, and requiring ambient or eventually low air pressure, which can be achieved with very economic pre-pumps and sealed boxes. In the next step, we will extend this result to higher pulse energies of several tens of mJ (our laser system can provide up to 50 mJ at 10 kHz) and stronger spectral broadening by slightly reducing the air pressure, chirping the input pulses and using optics with higher damage threshold. According to our calculations, this could be achieved with very compact and simple cells such as the one presented here.

**Funding:** Deutsche Forschungsgemeinschaft (390677874). We acknowledge support by the DFG Open Access Publication Funds of the Ruhr-Universität Bochum. Funded by the Deutsche Forschungsgemeinschaft (DFG, German Research Foundation) under Germanys Excellence Strategy – EXC-2033 – Projektnummer 390677874 - RESOLV. This publication was funded by Mercator Research Center Ruhr GmbH within the project "Towards an UARuhr Ultrafast Laser Science Center: Tailored fs-XUV Beamline for Photoemission Spectroscopy".

**Disclosures:** The authors declare no conflicts of interest.

**Data availability:** Data underlying the results presented in this paper are not publicly available at this time but may be obtained from the authors upon reasonable request.